\documentclass{article}
\usepackage{amssymb, amsmath}
\numberwithin{equation}{section}
\newtheorem{definition}{Definition}[section]
\newtheorem{theorem}[definition]{Theorem}

\newtheorem{lemma}[definition]{Lemma}

\def\min{{\rm min}}

\begin{document}

\title{The plane symmetric Einstein-dust system with positive cosmological constant}

\author{S. B. TCHAPNDA\footnote{On leave from Mathematics Department, University of Yaounde I, Box 812 Yaounde Cameroon}\\
\it Max Planck Institute for Gravitational Physics\\
\it Albert Einstein Institute\\
\it Am M\"uhlenberg 1, D-14476 Golm, Germany\\
\it \texttt{tchapnda@aei.mpg.de}}

\date{}
\maketitle{}
\begin{abstract}
The Einstein equations with a positive cosmological constant are
coupled to the pressureless perfect fluid matter in plane
symmetry. Under suitable restrictions on the initial data, the
resulting Einstein-dust system is proved to have a global
classical solution in the future time direction. Some late time asymptotic properties are obtained as well. 
\end{abstract}
\section{Introduction}
In \cite{tchapnda} plane symmetric solutions of the
Einstein-Vlasov system with positive cosmological constant were
investigated. It was shown that a spacetime of this type which is
initially expanding exists globally in the future when expressed
in an areal time coordinate $t$ and information was obtained about
its asymptotics for $t \to \infty$. It is future geodesically
complete and resembles the de Sitter solution at late times.
Information is obtained on the decay rates of the components of
the energy-momentum tensor.

This paper is concerned with the question, to what extent
analogues of these results for the Einstein-Vlasov system hold in
the case of the Einstein-dust system. There is an issue which has
to be addressed right at the start. This is that of shell-crossing
singularities. The intuitive idea behind this concept, as explained in \cite{isenberg} and \cite{rendall3}, is the following.
A shell of dust particles which are related to each other by the symmetry
of the spacetime moves in a coherent way. If two of these shells collide
then the intermediate shells are trapped between them, so that the matter density is forced to blow up. For more information about shell-crossing singularities see for instance \cite{nolan} and references therein.
For general plane symmetric solutions of the
Einstein-dust system it must be expected that shell-crossing
singularities develop, even from smooth initial data. For that
reason global classical solutions cannot be expected to exist. In
order to have a global existence theorem in the framework of
classical solutions it is therefore likely to be necessary to make some kind of smallness assumption on the initial data. There is another fact which is essential in the following. This is the presence of a positive cosmological constant, known as a mechanism leading to solutions of the Einstein equations with exponential expansion. Results related to this idea have been obtained in \cite{tchapnda} and \cite{rendall1}. They are used in this paper to guess the decay rates on the geometric and matter quantities providing a basis to the bootstrap argument for the proof of the main result (Theorem \ref{t1}).

The rest of the paper is organised as follows. In section 2 the equations are derived and a local existence theorem is obtained for the corresponding Cauchy problem. In section 3 the solution is shown to exist globally in the future time direction and its late-time asymptotic behaviour is investigated, provided some restrictions on the initial datum.
\section{Preliminaries}
\subsection{The Einstein equations}
Let $(M,g)$ be a spacetime, where the manifold is assumed to be $M=I\times \mathbb{T}^3$, $I$ is a real interval and $\mathbb{T}^3=S^1 \times S^1 \times S^1$ is the three-torus. The metric $g$ and the matter fields are required to be invariant under the action of the Euclidean group $E_2$ on the universal cover. It is also required that the spacetime has an $E_2$-invariant Cauchy surface of constant areal time. In such conditions the metric is assumed to have the form
\begin{equation} \label{1.1}
  ds^2 = -e^{2\eta(t,x)}dt^2 + e^{2\lambda(t,x)}dx^2 + t^2
  (dy^2 + dz^{2}),
\end{equation}
where $t>0$, $\eta$ and $\lambda$ are periodic in $x$, and $y$ and $z$ range in $[0,2\pi]$.

The Einstein equations read
\begin{align}\label{einstein}
G^{\alpha\beta} + \Lambda g^{\alpha\beta}  = 8 \pi
T^{\alpha\beta},
\end{align}
where $G^{\alpha\beta}$ is the Einstein tensor, $T^{\alpha\beta}$ the energy-momentum tensor and $\Lambda$ is the cosmological constant we assume to be positive. We introduce the notation $\rho=e^{2\eta}T^{00}$, $j=e^{\lambda+\eta}T^{01}$ and $S=e^{2\lambda}T^{11}$.

After computations in plane symmetry with the previous coordinates considerations, we obtain from (\ref{einstein}) the following equations where the subscripts $t$ and $x$ refer to partial derivatives with respect to $t$ and $x$ respectively:
\begin{equation} \label{1.4}
e^{-2\eta} (2t\lambda_t+1) - \Lambda t^{2} = 8 \pi t^{2}\rho
\end{equation}
\begin{equation} \label{1.5}
e^{-2\eta} (2t\eta_t-1)+ \Lambda t^{2} = 8 \pi t^{2}S
\end{equation}
\begin{equation} \label{1.6}
\eta_x = -4 \pi t e^{\lambda+\eta}j
\end{equation}
\begin{equation} \label{1.7}
e^{-2\lambda}\left(\eta_{xx} + \eta_x(\eta_x - \lambda_x)\right) -
e^{-2\eta}\left(\lambda_{tt}+(\lambda_t-
\eta_t)(\lambda_t+\frac{1}{t})\right) + \Lambda  = 0.
\end{equation}

\subsection{The equations for dust}
We consider a pressureless perfect fluid with energy density $\mu:=\mu(t,x)>0$ and $4$-velocity $U^\alpha$. The latter is normalized to be of unit length $U^\alpha U_\alpha=-1.$ The plane symmetry allows us to set $U^\alpha:=\xi(e^{-\eta},e^{-\lambda}u,0,0)$ where $\xi=(1-u^2)^{-1/2}$ is the relativistic factor, $u:=u(t,x)$ being the scalar velocity that satisfies $|u|<1$. The energy momentum tensor for a pressureless perfect fluid is
\[
T^{\alpha\beta}=\mu U^\alpha U^\beta,
\]
that is
\begin{align*}
T^{00}&=e^{-2\eta}\frac{\mu}{1-u^2}=:e^{-2\eta}\rho\\
T^{01}&=e^{-\lambda-\eta}\frac{\mu u}{1-u^2}=:e^{-\lambda-\eta}j\\
T^{11}&=e^{-2\lambda}\frac{\mu u^2}{1-u^2}=:e^{-2\lambda}S,
\end{align*}
the other components being zero.

The equations for dust are given by 
\begin{equation}\label{euler}
\nabla_\alpha T^{\alpha\beta}=0.
\end{equation}
The components $\nabla_\alpha T^{\alpha 2}$ and $\nabla_\alpha T^{\alpha 3}$ vanish identically. Computing the remaining two components gives
\begin{equation}\label{1.8}
(e^\lambda\rho)_t+(e^{\eta}j)_x=-\lambda_te^\lambda S-\eta_x e^{\eta}j-\frac{2}{t}e^\lambda\rho
\end{equation}
\begin{equation}\label{1.9}
(e^\lambda j)_t+(e^{\eta}S)_x=-\lambda_te^\lambda j-\eta_x e^{\eta}\rho-\frac{2}{t}e^\lambda j,
\end{equation}
and expressing the equations (\ref{1.8})-(\ref{1.9}) in terms of the variables $\mu$ and $u$ gives
\begin{align}
(1-u^2)&[\mu_t+ue^{\eta-\lambda}\mu_x]+\mu[2uu_t+(1+u^2)e^{\eta-\lambda}u_x]\nonumber\\&=-\mu(1-u^2)[(1+u^2)\lambda_t+2t^{-1}+2u\eta_xe^{\eta-\lambda}]\label{1.11}\\
(1-u^2)u&[\mu_t+ue^{\eta-\lambda}\mu_x]+\mu[(1+u^2)u_t+2ue^{\eta-\lambda}u_x]\nonumber\\&=-\mu(1-u^2)[2u(\lambda_t+t^{-1})+(1+u^2)\eta_xe^{\eta-\lambda}]. \label{1.12}
\end{align}
Adding and subtracting (\ref{1.11}) and (\ref{1.12}) yields
\begin{align}
(1-u)[\mu_t+ue^{\eta-\lambda}\mu_x]&+\mu[u_t+e^{\eta-\lambda}u_x]\nonumber\\&=-\mu[(1-u^2)(\lambda_t+\eta_xe^{\eta-\lambda})+2(1-u)t^{-1}]\label{1.13}\\
(1+u)[\mu_t+ue^{\eta-\lambda}\mu_x]&+\mu[-u_t+e^{\eta-\lambda}u_x]\nonumber\\&=-(1+u)\mu[(1-u)\lambda_t+2t^{-1}-(1-u)\eta_xe^{\eta-\lambda}], \label{1.14}
\end{align}
and the linear combinations $(\ref{1.13})+(\ref{1.14})$ and $(1+u)(\ref{1.13})+(u-1)(\ref{1.14})$ lead to
\begin{align}
D\mu&=-\mu\left[(1-u^2)\lambda_t-2t^{-1}+e^{\eta-\lambda}u_x\right]\label{1.15}\\
Du&=(1-u^2)\left[-u\lambda_t-e^{\eta-\lambda}\eta_x\right], \label{1.16}
\end{align}
where the derivative is
\begin{equation*}
D:=\partial_t+ue^{\eta-\lambda}\partial_x.
\end{equation*}
This is called the characteristic derivative and we denote the corresponding integral curve by $(t,\gamma)$. This means that this curve satifies the differential equation \\$\gamma_t=ue^{\eta-\lambda}$, and on the curve $D=\frac{d}{dt}$  so that we can for instance rewrite (\ref{1.15}) in the form
\begin{equation*}
\frac{d}{dt}\mu(t,\gamma(t))=-\mu\left[(1-u^2)\lambda_t-2t^{-1}+e^{\eta-\lambda}u_x\right](t,\gamma(t)).
\end{equation*}
\subsection{The Cauchy problem and local existence}
The object of our study is the plane symmetric Eintein-dust system (\ref{1.4})-(\ref{1.7}), (\ref{1.15})-(\ref{1.16}) with unknowns $\lambda$, $\eta$, $\mu$ and $u$. The initial data are prescribed at some time $t_0>0$. To analyze the solutions of these equations, the first step is to obtain a local existence theorem. The method, which has been used for instance in \cite{rein} for the Einstein-Vlasov system in plane symmetry, consists on constructing an iteration and proving its convergence. In the present investigation we follow another approach, which is described in \cite{rendall3}. There are several steps and it is not convenient to keep track of the differentiability in the process. For this reason only the case of $C^\infty$ initial data will be treated in this paper. There is a general (without symmetry) local existence theorem for the Einstein-dust system by Choquet-Bruhat \cite{choquet}.\\
In fact in that reference the author writes the Einstein-dust system (\ref{einstein})-(\ref{euler}) in the equivalent form
\begin{align}
u^\nu\nabla_\nu G_{\alpha\beta}&=-8\pi\mu u_\alpha u_\beta\nabla_\nu u^\nu\\
u^\nu \nabla_\nu u_\beta&=0\\
\nabla_\nu(\mu u^\nu)&=0,
\end{align}
in order to prove that the  equations are hyperbolic in the Leray sense. Then by the Leray-Dionne theory \cite{leray}-\cite{dionne} this implies existence and uniqueness of the solution to the Cauchy problem for Einstein-dust equations. For more details see also \cite{friedrich}.\\
We can apply this general result to the case with symmetry. Consider a plane symmetric compact $C^\infty$ initial data for the Cauchy problem. Then the symmetry is inherited by the corresponding solutions. (See \cite{helmut}, section 5.6 for a discussion of this.) Next areal coordinates can be introduced in the spacetime. The conclusion is a local existence and uniqueness theorem for the plane symmetric Einstein-dust system. The solution, like the initial data, is $C^\infty$.

\section{Global existence to the future and asymptotics}

This section is concerned with the main result of this paper. We first prove two lemmas dealing with bounds on the unknowns as well as all their derivatives. 

In what follows $C$ will denote a positive constant estimating functions that are uniformly bounded, and it may change from line to line.\\ Let $P_n(t):=||\partial_x^n\lambda(t)||_\infty+||\partial_x^n\mu(t)||_\infty+||\partial_x^nw(t)||_\infty$
\begin{lemma}\label{l1}
Consider a $C^\infty$ plane symmetric solution of Einstein-dust system on a time interval $[0,T)$ such that the following estimates hold, where $w:=e^{-\lambda}u_x$:
\begin{equation}\label{hd02}
|\lambda_t-t^{-1}|\leq Ct^{-4},\ |\Lambda t^2e^{2\eta}-3|\leq Ct^{-3},\ |\lambda_{tx}|\leq Ct^{-4}
\end{equation}
\begin{equation}\label{hd03}
|\eta_x|\leq Ct^{-3},\ |\eta_{xx}|\leq Ct^{-3},\ \mu\leq Ct^{-3}
\end{equation}
\begin{equation}\label{hd04}
|u|\leq Ct^{-1},\ |\mu_x|\leq Ct^{-3},\ |u_x|\leq Ct^{-1},\ |w_x|\leq Ct^{-2}.
\end{equation}
If all derivatives with respect to $x$ of order up to $n$ of the quantities $\lambda$,  $\mu$, $w$,  $\eta$ and $u$ are bounded then all derivatives with respect to $x$ of order up to $n+1$ of the same quantities are bounded.
\end{lemma}
{\bf Proof.}  Note that from the hypotheses of the lemma it follows that the quantities $\lambda$,  $\mu$, $w$,  $\eta$ and $u$, as well as all their first order derivatives are bounded on $[0,T)$.

By definition $w=e^{-\lambda}u_x$, which implies
\begin{equation}\label{hd00}
u_{xx}=e^\lambda w_x+\lambda_xu_x,
\end{equation}
and from (\ref{1.6}),
\begin{equation}\label{hd01}
\eta_{xx}=-4\pi te^{\eta+\lambda}j_x+\eta_x(\eta_x+\lambda_x).
\end{equation}
Differentiating $u_{xx}$ and $\eta_{xx}$ $n-1$ times with respect to $x$,  the boundedness of $\partial_x^{n+1}\eta$ and $\partial_x^{n+1}u$ follows immediately from the hypotheses of the lemma.\\
 The expression for $\lambda_{tx}$ is
\begin{equation}\label{hd1}
\lambda_{tx}=\eta_xe^{2\eta}(\Lambda t+8\pi t\rho)+4\pi t\rho_{x}e^{2\eta}.
\end{equation}
Differentiating this $n$ times with respect to $x$ gives a linear equation for $\partial_t(\partial_x^{n+1}\lambda)$ with coefficients which are known to be bounded, except for terms involving $\partial_x^{n+1}\rho$. But the latter can be estimated in terms of $\partial_x^{n+1}\mu$ so that the following holds 
\begin{equation}\label{hd2}
|\partial_t(\partial_x^{n+1}\lambda)|\leq C(1+|\partial_x^{n+1}\mu|).
\end{equation}
 Integrating this in time implies that
\begin{equation}\label{hd3}
||\partial_x^{n+1}\lambda(t)||_\infty\leq
||\partial_x^{n+1}\lambda(t_0)||_\infty+C\int_{t_0}^t(1+P_{n+1}(s))\ ds.
\end{equation}
Now recall that
\begin{align}\label{hd4}
D\mu_x&=-\mu_x[(1-u^2)\lambda_t-2t^{-1}+2e^{\eta-\lambda}u_x+u(\eta_x-\lambda_x)e^{\eta-\lambda}]\nonumber\\
&-\mu[-2uu_x\lambda_t+(1-u^2)\lambda_{tx}+(\eta_x-\lambda_x)e^{\eta-\lambda}u_x+e^{\eta-\lambda}u_{xx}],
\end{align}
and
\begin{align}\label{hd5}
D&(w_x)=-\Lambda tu(1-u^2)(\eta_{xx}+2\eta^2_x-\lambda_x\eta_x)e^{2\eta-\lambda}-2\eta_xwe^\eta[w+t\Lambda(1-3u^2)e^\eta]\nonumber\\&+3t^{-1}uu_xw(\Lambda t^2e^{2\eta}-1)-uw[\eta_{xx}+\eta_x(\eta_x-\lambda_x)]e^{\eta-\lambda}-\lambda_{tx}w\nonumber\\&-\left[\lambda_t+[2u_x+u(2\eta_x-\lambda_x)]e^{\eta-\lambda}+\frac{t^{-1}(1-3u^2)}{2}(\Lambda t^2e^{2\eta}-1)\right]w_x.
\end{align}
Differentiating (\ref{hd4}) $n$ times with respect to $x$ and using (\ref{hd01}) and (\ref{hd1}) shows that $D(\partial_x^{n+1}\mu)$ depends linearly on
$\partial_x^{n+1}\lambda$, $\partial_x^{n+1}\mu$ and $\partial_x^{n+2}u$ with bounded
coefficients. But $\partial_x^{n+2}u$ can be estimated in terms of $\partial_x^{n+1}w$ and $\partial_x^{n+1}\lambda$. It then follows that
\begin{equation}\label{hd6}
|D(\partial_x^{n+1}\mu)|\leq C(1+|\partial_x^{n+1}\mu|+|\partial_x^{n+1}\lambda|+|\partial_x^{n+1}w|),
\end{equation}
and integrating this along the characteristic $\gamma$ implies
that
\begin{equation}\label{hd7}
||\partial_x^{n+1}\mu(t)||_\infty\leq||\partial_x^{n+1}\mu(t_0)||_\infty+
C\int_{t_0}^t(1+P_{n+1}(s))\ ds.
\end{equation}
Likewise taking the $x$-derivative $n$ times in (\ref{hd5}) leads to
\begin{equation}\label{hd8}
|D(\partial_x^{n+1}w)|\leq C(1+|\partial_x^{n+1}\mu|+|\partial_x^{n+1}\lambda|+|\partial_x^{n+1}w|),
\end{equation}
and integration along $\gamma$ implies
\begin{equation}\label{hd9}
||\partial_x^{n+1}w(t)||_\infty \leq||\partial_x^{n+1}w(t_0)||_\infty+
C\int_{t_0}^t(1+P_{n+1}(s))\ ds.
\end{equation}
Putting (\ref{hd3}), (\ref{hd7}) and (\ref{hd9}) together implies
\begin{equation}\label{hd10}
P_{n+1}(t)\leq P_{n+1}(t_0)+ C\int_{t_0}^t(1+P_{n+1}(s))ds.
\end{equation}
By Gronwall's inequality it follows that $P_{n+1}$ is bounded and
thus so are $\partial_x^{n+1}\lambda$, $\partial_x^{n+1}\mu$ and $\partial_x^{n+1}w$. This
completes the proof of the lemma. \ \  $\Box$

\begin{lemma}\label{l2}
If the hypotheses of Lemma \ref{l1} are satisfied and if all derivatives of the quantities $\lambda$, $\eta$, $\mu$ and $u$ of the form $\partial_t^k\partial_x^n$ with $n$ arbitrary and $k\leq m$ are bounded then the derivatives of the form $\partial_t^{m+1}\partial_x^n$ of the same quantities are bounded
\end{lemma}
{\bf Proof.} From the evolution equations we have
\begin{align}
\lambda_t&=\frac{1}{2}(\Lambda te^{2\eta}-t^{-1})+4\pi t e^{2\eta}\rho\label{hd11}\\
\eta_t&=\frac{1}{2}(t^{-1}-\Lambda te^{2\eta})+4\pi t e^{2\eta}S\label{hd12}\\
\mu_t&=-ue^{\eta-\lambda}\mu_x-\mu\left[(1-u^2)\lambda_t-2t^{-1}+e^{\eta-\lambda}u_x\right]\label{hd13}\\
u_t&=-ue^{\eta-\lambda}u_x+(1-u^2)\left[-u\lambda_t-e^{\eta-\lambda}\eta_x\right]. \label{hd14}
\end{align}
Differentiating (\ref{hd11})-(\ref{hd14}) $n$ times with respect to $x$ and $m$ times with respect to $t$ allows $\partial_t^{m+1}\partial_x^n \lambda$, $\partial_t^{m+1}\partial_x^n \eta$, $\partial_t^{m+1}\partial_x^n \mu$ and $\partial_t^{m+1}\partial_x^n u$ to be bounded.\ \ \ \ \ \ $\Box$

We can now prove the main result of the present investigation.
\begin{theorem}\label{t1}
Consider any $C^\infty$ solution of Einstein-dust system with positive cosmological constant in plane symmetry written in areal coordinates with $C^\infty$ initial data. Let $\delta$ be a positive constant and suppose the following inequalities hold:
\begin{equation}\label{bo1}
|\lambda_t(t_0)-t_{0}^{-1}|\leq \delta,\ |\Lambda t_{0}^2e^{2\eta(t_0)}-3|\leq \delta,\ |\eta_x(t_0)|\leq \delta,\ |\lambda_{tx}(t_0)|\leq \delta,
\end{equation}
\begin{equation}\label{bo2}
|\eta_{xx}(t_0)|\leq \delta,\  \mu(t_0)\leq \delta,\ |u(t_0)|\leq \delta,\ |\mu_x(t_0)|\leq \delta,\ |u_x(t_0)|\leq \delta,\ |w_x(t_0)|\leq \delta.
\end{equation}
Then if $\delta$ is sufficiently small, the corresponding solution exists on $[t_0,\infty)$. Moreover, for this solution the following properties hold at late times:
\begin{equation}\label{bo3}
|\lambda_t-t^{-1}|=O(t^{-4}),\ |\Lambda t^2e^{2\eta}-3|=O(t^{-3}),\ |\eta_x|=O(t^{-3}),
\end{equation}
\begin{equation}\label{bo4}
|\eta_{xx}|=O(t^{-3}),\ |\lambda_{tx}|=O(t^{-4}),\ \mu=O(t^{-3}),
\end{equation}
\begin{equation}\label{bo5}
|\mu_x|=O(t^{-3}),\ |u|=O(t^{-1}),\ |u_x|=O(t^{-1}),\ |w_x|=O(t^{-2}).
\end{equation}
\end{theorem}
{\bf Proof.} The proof proceeds by a bootstrap argument.

By continuity it follows from the hypothesis (\ref{bo1})-(\ref{bo2}) that
\begin{equation*}
|\lambda_t(t)-t^{-1}|\leq 2\delta,\ |\Lambda t^2e^{2\eta(t)}-3|\leq 2\delta,\ |\eta_x(t)|\leq 2\delta,\ |\eta_{xx}(t)|\leq 2\delta
\end{equation*}
\begin{equation*}
|\lambda_{tx}(t)|\leq 2\delta,\ \mu(t)\leq 2\delta,\ |u(t)|\leq 2\delta,\ |\mu_x|\leq 2\delta,\ |u_x|\leq 2\delta,\ |w_x|\leq 2\delta,
\end{equation*}
for $t$ close to $t_0$.

Let $C_1$ and $\varepsilon$ be constants for $0<C_1<1$ and $0<\varepsilon<1/2$. We can reduce $\delta$ if necessary so that $2\delta <C_1\min(t_{0}^{-4+\varepsilon},t_{0}^{-1+\varepsilon})$. Then there exists some time interval on which the solution of the Einstein-dust system exists and the following bootstrap assumption is satisfied
\begin{equation}\label{bo6}
|\lambda_t-t^{-1}|\leq C_1t^{-4+\varepsilon},\ |\Lambda t^2e^{2\eta}-3|\leq C_1t^{-3+\varepsilon},\ |\lambda_{tx}|\leq C_1t^{-4+\varepsilon}
\end{equation}
\begin{equation}\label{bo7}
|\eta_x|\leq C_1t^{-3+\varepsilon},\ |\eta_{xx}|\leq C_1t^{-3+\varepsilon},\ \mu\leq C_1t^{-3+\varepsilon}
\end{equation}
\begin{equation}\label{bo8}
|u|\leq C_1t^{-1+\varepsilon},\ |\mu_x|\leq C_1t^{-3+\varepsilon},\ |u_x|\leq C_1t^{-1+\varepsilon},\ |w_x|\leq C_1t^{-2+\varepsilon}.
\end{equation}
Consider the maximal interval $[t_0,t_*)$ on which the solution  of the full system (\ref{1.4})-(\ref{1.7}), (\ref{1.15})-(\ref{1.16}) exists and (\ref{bo6})-(\ref{bo8}) hold. Suppose $t_*$ is finite.

Putting inequalities (\ref{bo6})-(\ref{bo8}) into equations coming from the system (\ref{1.4})-(\ref{1.7}), (\ref{1.15})-(\ref{1.16}) allows new estimates to be derived. We first derive an estimate for $u$. For this purpose an evolution equation for $tu$ can be obtained from (\ref{1.16}), using the field equations (\ref{1.4})-(\ref{1.5}) involving $\lambda_t$ and $\eta_x$. The result is
\begin{equation}\label{bo9}
D(tu)=u^3-\frac{u}{2}(1-u^2)(\Lambda e^{2\eta}t^2-3).
\end{equation}
Using the bootstrap assumption on $|u|$ and $|\Lambda t^2e^{2\eta}-3|$, integrating the resulting inequality along the integral curve $\gamma$ and keeping the worst powers it follows from (\ref{bo9}) that
\begin{equation}\label{bo10}
|u(t)|\leq[t_0|u(t_0)|+C_1^3+C_1^2]t^{-1}=:C_2t^{-1}.
\end{equation}
Next we derive an estimate for $\mu$. Using (\ref{1.15}) an evolution equation for $t^3\mu$ follows:
\begin{equation}\label{bo11}
D(t^3\mu)=t^3\mu\left[t^{-1}u^2-(1-u^2)(\lambda_t-t^{-1})-e^{\eta-\lambda}u_x\right].
\end{equation}
An estimate for $e^{2\eta}$, $e^{\eta-\lambda}$ and $e^{\eta+\lambda}$ will be also required.
\begin{align}
e^{2\eta}&=\Lambda^{-1}t^{-2}(\Lambda t^{2}e^{2\eta})\notag \\
&\leq \Lambda^{-1}t^{-2}[(\Lambda t^{2}e^{2\eta}-3)+3]\notag\\
&\leq 3\Lambda^{-1}t^{-2}+C_1 \Lambda^{-1}t^{-5+\varepsilon}.\label{bo12}
\end{align}
On the other hand by assumption $|\lambda_t-t^{-1}|\leq C_1 t^{-4+\varepsilon}$, and integrating this in time implies that $e^{-\lambda}\leq e^{C_1-\lambda(t_0)}t_0t^{-1}$ and $e^{\lambda}\leq e^{C_1+\lambda(t_0)}t_0^{-1}t$. Thus
\begin{equation}\label{bo13}
e^{\eta-\lambda}\leq\Lambda^{-1/2}(\sqrt3+C_1^{1/2})t_0e^{C_1-\lambda(t_0)}t^{-2},\ {\rm and} \ e^{\eta+\lambda}\leq\Lambda^{-1/2}(\sqrt3+C_1^{1/2})t_0^{-1}e^{C_1+\lambda(t_0)}.
\end{equation}
Using this, the bootstrap assumption, integration along $\gamma$ and keeping the worst powers it follows from (\ref{bo11}) that
\begin{equation}\label{bo14}
\mu\leq\left[t_0^3\mu+C_1C_2+C_1^2+C_1\Lambda^{-1/2}(\sqrt3+C_1^{1/2})t_0e^{C_1-\lambda(t_0)}\right]t^{-3}=:C_3t^{-3}.
\end{equation}
Now estimates for $|\Lambda t^2e^{2\eta}-3|$ and $|t\lambda_t-1|$ will be derived. Estimates for the matter quantities $S$ and $\rho$ are needed for this purpose.
From the definition of $S$ and $\rho$ and the estimates for $\mu$ and $u$ obtained above, we obtain
\begin{align}\label{bo15}
\rho\leq\frac{C_3}{1-C_2}t^{-3},\ S\leq\frac{C_3C_2^2}{1-C_2}t^{-5}.
\end{align}
From (\ref{1.5}) we have
\begin{equation*}
\partial_t[-\frac{1}{3}t e^{-2\eta}(\Lambda t^2e^{2\eta}-3)]=-8\pi t^2 S,
\end{equation*}
so that using (\ref{bo12}), (\ref{bo15}), integration and keeping the worst powers gives
\begin{align}\label{bo16}
|\Lambda& t^2e^{2\eta}-3|\nonumber\\&\leq[3\Lambda^{-1}+C_1 \Lambda^{-1}]\left[t_0 |\Lambda t_{0}^{2}-3e^{-2\mu(t_0)}|+\frac{24\pi C_3C_2^2}{1-C_2} \right]t^{-3}\nonumber\\&=:C_4t^{-3}.
\end{align}
From (\ref{1.4}) we have
\begin{equation*}
t\lambda_t-1=\frac{1}{2}(\Lambda e^{2\eta}t^2-3)+4\pi t^2 e^{2\eta}\rho,
\end{equation*}
and using (\ref{bo12}), (\ref{bo15}), (\ref{bo16}) and keeping the worst powers yields
\begin{equation}\label{bo17}
|t\lambda_t-1|\leq \left[C_4+4\pi\Lambda^{-1}(3+C_1)\frac{C_3}{1-C_2}\right]t^{-3}=:C_5t^{-3}.
\end{equation}
An estimate for $\eta_x$ will be derived. Recalling that $$\eta_x=-4\pi te^{\lambda+\eta}j \ {\rm and} \ j=\frac{\mu u}{1-u^2},$$ it follows from (\ref{bo10}) and (\ref{bo13})-(\ref{bo14}) that
\begin{align}\label{bo18}
|\eta_x|&\leq 4\pi \Lambda^{-1/2}(C_{1}^{1/2}+\sqrt{3})e^{C_1+\lambda(t_0)}\frac{C_2C_3}{1-C_2}t^{-3}\nonumber\\&=:C_6t^{-3}.
\end{align}
The following are estimates for $u_x$ and $\mu_x$ which will be needed in order to get better estimates for $\lambda_{tx}$ and $\eta_{xx}$.\\
An evolution equation for $tu_x$ can be derived by taking the $x$-derivative in (\ref{bo9}):
\begin{align}\label{bo19}
D(tu_x)=(3u^2-1)u_x&(t\lambda_t-1)+3u_xu^2+t(\eta_x+\lambda_x)uu_xe^{\eta-\lambda}-tu^2_xe^{\eta-\lambda}\nonumber\\&-t(1-u^2)[u\lambda_{tx}+(\eta_x-\lambda_x)\eta_xe^{\eta-\lambda}+\eta_{xx}e^{\eta-\lambda}],
\end{align}
and using the bootstrap assumption, some estimates obtained above as well as integration along $\gamma$ leads to
\begin{align}\label{bo20}
|u_x|&\leq\left[t_0|u_x(t_0)|+4C_1^2+4t_0C_1\Lambda^{-1/2}(\sqrt3+C_1^{1/2})e^{C_1-\lambda(t_0)}(2C_1+|\lambda_x(t_0)|)\right]t^{-1}\nonumber\\&=:C_7t^{-1}.
\end{align}
Now taking the $x$-derivative in (\ref{1.15}) allows us to obtain an evolution equation for $t^3\mu_x$:
\begin{align}\label{bo21}
D(t^3\mu_x)&=-t^2(t\lambda_t-1)[(1-u^2)\mu_x-2\mu uu_x]-2t^3\mu_xu_xe^{\eta-\lambda}\nonumber\\&-t^3(\eta_x-\lambda_x)u\mu_xe^{\eta-\lambda}+t^2(u^2\mu_x+2\mu uu_x)-t^3(1-u^2)\mu \lambda_{tx}\nonumber\\&-t^3(\eta_x-\lambda_x)\mu u_xe^{\eta-\lambda}-t^3\mu u_{xx}e^{\eta-\lambda}.
\end{align}
So far we know how to estimate all the terms on the right hand side in (\ref{bo21}) except the last one in $u_{xx}$ which needs to be worked out carefully. We can obtain an evolution equation for $u_{xx}$ by taking the $x$-derivative of $Du_x$. This does not give a satisfactory result because  terms containing $\lambda_{xx}$ occur, and we do not know how to estimate them. This difficulty can be overcome if we rather take the $x$-derivative of $D(e^{-\lambda}u_x)$, which will lead to an estimate for $w_x$ (where $w=e^{-\lambda}u_x$). An estimate for $u_{xx}$ will then be deduced from the following relation which is obtained by differentiating the equality $w=e^{-\lambda}u_x$ in $x$:
\begin{equation}\label{bo22}
u_{xx}=e^\lambda w_x+\lambda_xu_x.
\end{equation}
Note that the factor $e^{-\lambda}$ allows us to eliminate bad terms such as $\lambda_{xx}$. This device has been used in another context \cite{tegankong}.\\
The term $w_x$ will now be estimated. An evolution equation for $w$ is
\begin{align*}
Dw&=-\lambda_tw-w^2e^\eta -u\eta_xwe^{\eta-\lambda}-t^{-1}w(1-3u^2)\\&-\frac{t^{-1}w(1-3u^2)}{2}(\Lambda t^2e^{2\eta}-3)-\Lambda t\eta_xu(1-u^2)e^{2\eta-\lambda},
\end{align*}
so that differentiating this in $x$ implies
\begin{align}\label{bo23}
D&(t^2w_x)=-\Lambda t^3u(1-u^2)(\eta_{xx}+2\eta^2_x-\lambda_x\eta_x)e^{2\eta-\lambda}-2t^2\eta_xwe^\eta[w+t\Lambda(1-3u^2)e^\eta]\nonumber\\&+6tuu_xw+3tuu_xw(\Lambda t^2e^{2\eta}-3)-t^2uw(\eta_{xx}+\eta_x(\eta_x-\lambda_x))e^{\eta-\lambda}-t^2\lambda_{tx}w\nonumber\\&-\left[t(t\lambda_t-1)-3tu^2+t^2[2u_x+u(2\eta_x-\lambda_x)]e^{\eta-\lambda}+\frac{t(1-3u^2)}{2}(\Lambda t^2e^{2\eta}-3)\right]w_x.
\end{align}
We use the bootstrap assumption, some estimates obtained above and integration along $\gamma$ to get
\begin{align}\label{bo24}
|w_x|&\leq[t_0^2|w_x(t_0)|+4C_7C_1t_0e^{C_1-\lambda(t_0)}(1+|\lambda_x(t_0)|)\nonumber\\&+4C_1^2(1+\Lambda^{-1/2})(1+|\lambda_x(t_0)|)]t^{-2}\nonumber\\&=:C_8t^{-2}.
\end{align}
It then follows from (\ref{bo22}) that
\begin{align}\label{bo25}
|u_{xx}|\leq\left[C_8t_0^{-1}e^{C_1+\lambda(t_0)}+C_7(C_1+|\lambda_x(t_0)|)\right]t^{-1}=:C_{u_{xx}}t^{-1}.
\end{align}
This together with (\ref{bo21}) imply that
\begin{align}\label{bo26}
|\mu_{x}|&\leq[t_0^{3}|\mu_x(t_0)|+7C_1^2\nonumber\\&+(3C_1+3C_{u_{xx}}+|\lambda_x(t_0)|)(\sqrt3+C_1^{1/2})\Lambda^{-1/2}t_0C_1e^{C_1-\lambda(t_0)}]t^{-3}\nonumber\\&=:C_{9}t^{-3}.
\end{align}
Estimates for $\rho_x$ and $j_x$ are required in order to derive estimates for $\lambda_{tx}$ and $\eta_{xx}$. We have
 $$\rho_x=\frac{\mu_x}{1-u^2}+\frac{2uu_x\mu}{(1-u^2)^2};$$ using estimates established above gives
\begin{equation*}
|\rho_x|\leq\frac{C_{9}+2C_2C_3C_8}{(1-C_1)^2}t^{-3},
\end{equation*}
and recalling that $$\lambda_{tx}=\eta_x e^{2\eta}(\Lambda t+8\pi t \rho)+4\pi t \rho_x e^{2\eta}$$ it follows that
\begin{align}\label{bo27}
   |\lambda_{tx}|&\leq\left[\Lambda^{-1}C_6(3+C_1)\left(\Lambda+\frac{8\pi C_3}{1-C_1}+4\pi \frac{C_{9}+2C_2C_3C_8}{(1-C_1)^2}\right)\right]t^{-4}\nonumber\\
   &=:C_{10}t^{-4}.
\end{align}
Now $$j_x=\frac{\mu_xu+\mu u_x}{1-u^2}+\frac{2u^2u_x\mu}{(1-u^2)^2}$$ so that using estimates obtained above  gives
\begin{align}\label{bo28}
|j_x|\leq\frac{C_2C_{9}+C_3C_8(2C_2^2+1)}{(1-C_1)^2}t^{-4}=:C_{j_x}t^{-4}.
\end{align}
Taking the spatial derivative of (\ref{1.6}) gives
\begin{equation}\label{bo29}
\eta_{xx}=-4\pi t e^{\eta+\lambda}j_x+\eta_x(\eta_x+\lambda_x).
\end{equation}
It then follows that
\begin{align}\label{bo30}
|\eta_{xx}|&\leq \left[4\pi \Lambda^{-1/2}(C_{1}^{1/2}+\sqrt{3})e^{C_1+\lambda(t_0)}C_{j_x}+C_4(C_4+C_1+|\lambda_x(t_0)|)\right]t^{-3}\nonumber\\&=:C_{11}t^{-3}.
\end{align}
The constants $C_2-C_{11}$ appearing along the proof are all less
than or equal to $C \times (g(\delta)+C_1^2)$, with $C$ a positive
constant and $g(\delta)$ a positive function of $\delta$ tending
to $0$ as $\delta$ tends to $0$. Therefore it is always possible
to choose $C_1$ and $\delta$ small enough in such a way that $CC_1
\leq 1/2$ and $Cg(\delta)\leq C_1/2$, and so the constants
$C_2-C_{11}$ are all less than $C_1$. This together with lemmas \ref{l1}-\ref{l2} show that all derivatives of $\lambda$, $\eta$, $\mu$ and $u$ are bounded on $[0,t_*)$. This means that the solution can be extended to a time interval $[t_0,t_1)$  on which (\ref{bo6})-(\ref{bo8}) hold, with $t_1>t_*$. This contradicts the
maximality of the interval $[t_0,t_*)$. Therefore $t_*=\infty$ and
the proof of the theorem is complete.\ \ \ \ \ $\Box$

\vskip 10pt\noindent \textbf{Acknowledgement} : I am grateful to
A. D. Rendall for fruitful suggestions.

\end{document}